\documentclass[letterpaper]{ptephy_v1}

\usepackage{amsmath,amssymb}
\usepackage{mathtools}
\usepackage{url}
\usepackage{tikz}
\usetikzlibrary{decorations.pathmorphing,intersections,calc,arrows}
\tikzset{snake it/.style={decorate, decoration=snake}}

\begin{document}

\title{Feasibility of an experimental search for a resonance of a pion and a light nucleus}


\author{Hiroyuki Fujioka}
\affil{Department of Physics, Tokyo Institute of Technology, Meguro, Tokyo 152-8551, Japan \email{fujioka@phys.titech.ac.jp}}


\begin{abstract}%

A hypothesis is proposed herein, 
suggesting that a pion-nuclear resonance may be observed
in the $\alpha+d\to{}^6\mathrm{Li}(3.563)+\pi^0$ reaction.
The resonance has a $\pi NN\alpha$ structure, containing $\alpha NN$ and $\pi NN$ subsystems.
The former corresponds to the $A=6$ isotriplet
\{$^6\mathrm{He}_\text{g.s.}$, $^6\mathrm{Li}(3.563)$, $^6\mathrm{Be}_\text{g.s.}$\},
whereas the latter is  a hypothetical $NN$-decoupled dibaryon.
We propose an experiment to search for this resonance using the $^7\mathrm{Li}(p,d)$ reaction.
\end{abstract}

\subjectindex{D01, D15, D22, D25}

\maketitle

\section{Introduction}

Numerous investigations of the meson--nucleus bound states have been performed
to elucidate the meson--nucleus and meson--nucleon interactions~\cite{Metag2017}.
In particular, systems with a pseudoscalar meson and a few nucleons
have attracted considerable attention.
An example of such a system is the bound state of an antikaon ($\overline{K}$) and two nucleons ($N$), $\overline{K}NN$.
Two recent experiments at J-PARC independently reported
observations of the $\overline{K}NN$ bound state
in the $d(\pi^+,K^+)$ reaction~\cite{Ichikawa2015}
and the $^3\mathrm{He}(K^-,n)$ reaction~\cite{Yamaga2020}.
The possible existence of a bound or virtual state of the $\eta$-$^3\mathrm{He}$ system
was inferred from the sharp increase in the cross sections just above the threshold
observed in the $\gamma{}^3\mathrm{He}\to {}^3\mathrm{He}\eta$ reaction~\cite{Pheron2012}
and the $pd\to {}^3\mathrm{He}\eta$ reaction~\cite{Xie2017},
whereas no $\eta$-$^3\mathrm{He}$ bound state was observed
in the excitation function of the $dp\to pd\pi^0$ reaction~\cite{Adlarson2020}.
In addition, the existence of an $\eta' d$ bound state is theoretically predicted
in connection with the $U_A(1)$ anomaly in quantum chromodynamics,
and it may be investigated by using the $\gamma d\to \eta d$ reaction~\cite{Sekihara2018}.

A pionic system such as $\pi NN$ has been insufficiently examined in the past few decades.
Instead, 
a non-strange dibaryon, which is defined as a system with a baryon number of 2, 
has been extensively investigated~\cite{Clement2017}.
Recently, the existence of $d^*(2380)$ has been established using neutron--proton scattering~\cite{Adlarson2014}
as well as double pionic fusion such as $pn\to d\pi^0\pi^0$~\cite{Adlarson2011}. 
The isospin and spin-parity were determined to be $I(J^P)=0(3^+)$, 
and $d^*(2380)$ is also called a $\Delta\Delta$ dibaryon.
A candidate of an $N\Delta$ dibaryon with $I(J^P)=1(2^+)$ was also obtained through a partial-wave analysis~\cite{Oh1997}.
These dibaryons involve $p$-wave $\pi N$ interaction in a hadronic picture~\cite{Gal2014}.

In this letter, we revisit a $\pi NN$ system
in which any two constituents are in a relative $s$-wave.
In addition, we consider a six-nucleon system of a $\pi NN\alpha$,
which may be regarded as the resonance of a pion \{$\pi^+$, $\pi^0$, $\pi^-$\}
and the $A=6$ isotriplet \{$^6\mathrm{He}_\text{g.s.}$, $^6\mathrm{Li}(3.563)$, $^6\mathrm{Be}_\text{g.s.}$\} 
with a total charge of $+3$. 
A hypothesis is proposed, suggesting that a sharp structure observed in the excitation function 
of a pionic fusion reaction~\cite{Andersson2006} may be accounted for by the formation of a $\pi NN\alpha$ resonance.
To test the existence of the $\pi NN\alpha$ resonance, 
an experiment to search for the resonance in the $^7\mathrm{Li}(p,d)$ reaction~\cite{RCNP_E503} is planned.

The rest of this letter is organised as follows.
In Section 2, we briefly review some of the previous studies on the $\pi NN$ system.
Section 3 introduces the six-nucleon system $\pi NN\alpha$.
In Section 4, we discuss the pionic fusion experiment conducted at the CELSIUS storage ring facility in Uppsala, Sweden
and propose a novel interpretation of the experimental result.
The planned experiment and its feasibility are described in Section 5.
Finally, Section 6 concludes the letter.

\section{\unboldmath $\pi NN$ System With $I(J^P)=0(0^-)$}

We consider a $\pi NN$ system with $I(J^P)=0(0^-)$
in which any two particles are in a relative $s$-wave.
Then, the total isospin of a $\pi N$ subsystem must be 1/2.
It is well known that the $s$-wave $\pi N$ interaction is attractive in the isospin 1/2 channel
and repulsive in the isospin 3/2 channel~\cite{Ericson-Weise}.
Moreover, the two nucleons are in the spin-singlet state;
hence, the interaction between them is also attractive.

A possible existence of such a three-body system is supported
by two different theoretical calculations.
Garcilazo solved the Faddeev equations and obtained
a resonance at $2018\,\mathrm{MeV}$ ($42\,\mathrm{keV}$ above the $\pi NN$ threshold)
with a width of $1.75\,\mathrm{MeV}$~\cite{Garcilazo1997}. 
Ueda calculated the amplitude for this system and found a strong structure
at $4\text{--}5\,\mathrm{MeV}$ above the $\pi NN$ threshold amplitude,
with a width of $4\text{--}5\,\mathrm{MeV}$~\cite{Ueda1998}.

A notable feature of this three-body system is that it does not couple with a $NN$ system
because the quantum numbers $I(J^P)=0(0^-)$ cannot be realised owing to the Pauli principle.
This means that a partial-wave analysis for nucleon-nucleon scattering will not
provide any signature, unlike in the case of the $N\Delta$ and $\Delta\Delta$ dibaryons.
As the resonance will decay into $\pi^0pn$,
an experimental search is highly difficult.
To date, no experiment has been conducted
to search for the $\pi NN$ resonance in the vicinity of the $\pi NN$ threshold, predicted in Refs.~\cite{Garcilazo1997,Ueda1998}.
However, it should be noted
that the production of narrow dibaryons below $2100\,\mathrm{MeV}$
with the $\gamma d \to X\pi^0$ reaction was investigated,
which resulted in an upper limit of the cross section of 
$2\text{--}5\,\mathrm{\mu b}$~\cite{Siodlaczek2000},
although the sensitivity may not be sufficient to rule out the existence of such a narrow resonance.
\section{\unboldmath $\pi NN\alpha$ System With $I(J^P)=0(0^-)$}
When the $\pi NN$ system described in the previous section and a $^4\mathrm{He}$ nucleus (an $\alpha$ particle) are in a relative $s$-wave,
the entire system of $\pi NN\alpha$ will also have $I(J^P)=0(0^-)$.
To satisfy these quantum numbers, the subsystem $\alpha NN$ must have $I(J^P)=1(0^+)$
if this subsystem and the pion are in a relative $s$-wave.
As this subsystem has a mass number of 6 and an isospin of 1, it corresponds to the $A=6$ isotriplet
\{$^6\mathrm{He}_\text{g.s.}$, $^6\mathrm{Li}(3.563)$, $^6\mathrm{Be}_\text{g.s.}$\}.
$^6\mathrm{Li}(3.563)$ is the first excited $0^+$ state of $^6\mathrm{Li}$
and is an isobaric analogue state of $^6\mathrm{He}_\text{g.s.}$ and $^6\mathrm{Be}_\text{g.s.}$.

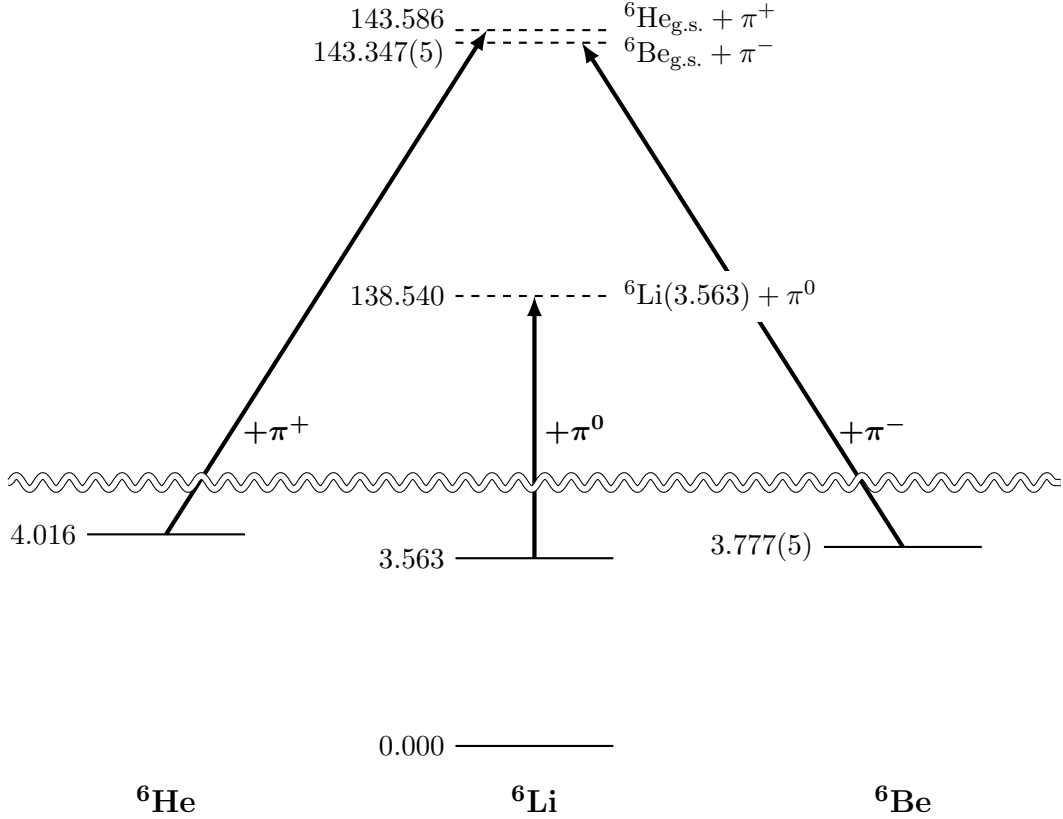
\begin{figure}[t]
    \newcommand{\len}{3} 
    \newcommand{\spa}{4}
    \def\LiEx{3.563}
    \def\HeEx{4.016}
    \def\BeEx{3.777}
    \def\offset{130}
    \def\chargedpi{139.570}
    \def\neutralpi{134.977}
    \centering
    \begin{tikzpicture}[scale=0.7]       
        \draw[thick](-0.5*\len,0)  node[left] {0.000} -- (0.5*\len,0) ;
        \draw[thick](-0.5*\len,\LiEx) node[left] {\LiEx}--(0.5*\len,\LiEx);
        \draw[thick](-1.5*\len-\spa,\HeEx) node[left] {\HeEx}--(-0.5*\len-\spa,\HeEx);
        \draw[thick](0.5*\len+\spa,\BeEx) node[left] {\BeEx(5)}--(1.5*\len+\spa,\BeEx);
        \node[font=\boldmath] at (0,-1) {\large $^6\mathrm{Li}$};
        \node[font=\boldmath] at (-\spa-\len,-1) {\large $^6\mathrm{He}$};
        \node[font=\boldmath] at (\spa+\len,-1) {\large $^6\mathrm{Be}$};

        \draw[thin, color=white, name path=L4](-2*\len-\spa,6)--(2*\len+\spa,6);

        \draw[ultra thick,->,>=latex,name path=L1] (-\len-\spa,\HeEx)-- (-0.3*\len,\HeEx+\chargedpi-\offset);
        \draw[ultra thick,->,>=latex,name path=L2] (0,\LiEx)-- (0,\LiEx+\neutralpi-\offset);
        \draw[ultra thick,->,>=latex,name path=L3] (\len+\spa,\BeEx)-- (0.3*\len,\BeEx+\chargedpi-\offset);

        \draw[thick,dashed](-0.5*\len,\LiEx+\neutralpi-\offset) node[left] {138.540}-- (0.5*\len,\LiEx+\neutralpi-\offset) node[right,fill=white]{$^6\mathrm{Li}(3.563)+\pi^0$};
        \draw[thick,dashed](-0.5*\len,\HeEx+\chargedpi-\offset) node[above=1.5mm, left] {143.586}-- (0.5*\len,\HeEx+\chargedpi-\offset) node[above=1.5mm, right]{$^6\mathrm{He}_\text{g.s.}+\pi^+$};
        \draw[thick,dashed](-0.5*\len,\BeEx+\chargedpi-\offset) node[below=1.5mm, left] {143.347(5)}-- (0.5*\len,\BeEx+\chargedpi-\offset) node[below=1.5mm, right]{$^6\mathrm{Be}_\text{g.s.}+\pi^-$};

        \draw[thin, double distance=2pt,snake it](-2*\len-\spa,5)--(2*\len+\spa,5);

        \path[name intersections={of= L1 and L4}];
        \node[right,font=\boldmath] at (intersection-1) {$+\pi^+$};
        \path[name intersections={of= L2 and L4}];
        \node[right,font=\boldmath] at (intersection-1) {$+\pi^0$};
        \path[name intersections={of= L3 and L4}];
        \node[right, font=\boldmath] at (intersection-1) {$+\pi^-$};
    \end{tikzpicture}
    \caption{Energy levels of the $A=6$ isotriplet (solid lines) and relevant thresholds (dashed lines). The energy relative to the ground state of $^6\mathrm{Li}$ is given in units of megaelectron volt.}
    \label{level}
\end{figure}

Owing to the mass difference in the $A=6$ triplet and that between the charged and neutral pions,
the relevant threshold energies for the $^6\mathrm{He}_\text{g.s.}+\pi^+$,
$^6\mathrm{Li}(3.563)+\pi^0$, and $^6\mathrm{Be} _\text{g.s.}+\pi^-$ channels are different, 
as illustrated in Fig.~\ref{level}.
Since the lowest threshold is that of the $^6\mathrm{Li}(3.563)+\pi^0$ channel,
the $\pi NN\alpha$ resonance will decay into $^6\mathrm{Li}(3.563)+\pi^0$.

To the best of the author’s knowledge, this six-nucleon system has never been proposed in the literature.
As in the $\pi NN$ system,
pion absorption into the $I=1$ di-nucleon $NN$ in the $s$-wave is forbidden in a strong interaction.
This means that the decay of the $\pi NN\alpha$ resonance into a non-pionic final state is suppressed.
Although a quantitative evaluation is needed, the decay width of the $\pi NN\alpha$ resonance may be narrow,
which is a remarkable feature from an experimental perspective.

In contrast to a deeply bound pionic state in a heavy nucleus~\cite{Yamazaki2012,Nishi2018}, 
of which the narrow absorption width is due to the repulsion in the strong interaction between a $\pi^-$ and a nucleus,
the strong interaction between the pion and the $A=6$ nucleus in the $\pi NN\alpha$ system is 
attractive owing to the pion--nucleon attraction in the $I=1/2$ channel.

\section{\unboldmath Hint of a Resonance Near the $^6\mathrm{Li}(3.563)+\pi^0$ Threshold}
The pionic fusion, i.e.~nuclear fusion resulting in coherent pion production, 
of a deuteron and an alpha particle was 
investigated at the CELSIUS storage ring facility in Uppsala, Sweden~\cite{Andersson2000,Andersson2006}.
The total cross section and the forward--backward asymmetry were supposed to be sensitive to
the cluster structure of $^6\mathrm{He}_\text{g.s.}$
and its isobaric analogue state of $^6\mathrm{Li}(3.563)$,
as an alpha core and a quasi-deuteron with $I=1$ and $J=0$.
For this purpose,
a $2s$ harmonic oscillator wave function was used to describe 
the relative motion of the alpha core and two-nucleon halo in the momentum space,
with the aim of determining the harmonic oscillator constant,
which can be converted into the point-proton charge radius of $^6\mathrm{He}_\text{g.s.}$ and $^6\mathrm{Li}(3.563)$.

\begin{figure}[t]
    \centering\includegraphics[width=15cm]{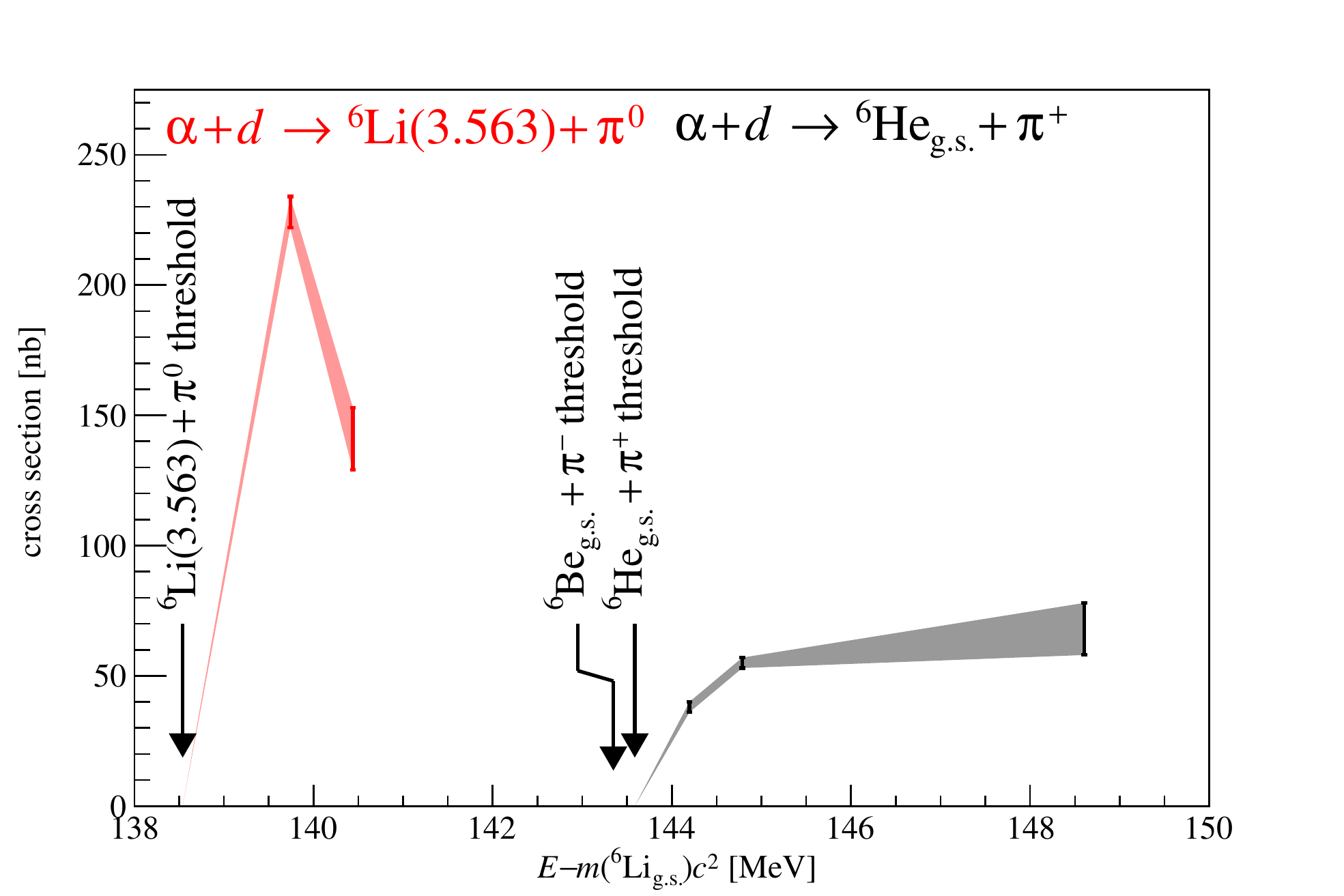}
    \caption{Cross sections for the $\alpha+d\to{}^6\mathrm{Li}(3.563)+\pi^0$ reaction (red) and the $\alpha+d\to{}^6\mathrm{He}_\text{g.s.}+\pi^+$ reaction (black) as a function of 
    the excitation energy, i.e. the centre-of-mass energy minus the rest energy of the ground state of $^6\mathrm{Li}$.
    The systematic error for a multiplicative factor in each reaction is not indicated. 
    The bands are drawn as a visual guide. 
    The data are taken from Ref.~\cite{Andersson2006}.}
    \label{CELSIUS}
\end{figure}

The cross sections for two pionic fusion reactions, 
$\alpha+d\to{}^6\mathrm{Li}(3.563)+\pi^0$ and $\alpha+d\to{}^6\mathrm{He}_\text{g.s.}+\pi^+$,
are illustrated in Fig.~\ref{CELSIUS}.
For the latter reaction, the cross sections were corrected for the Coulomb interaction in the final state.
By fitting the experimental results for the Coulomb-corrected cross section
and the asymmetry for the $\alpha+d\to{}^6\mathrm{He}_\text{g.s.}+\pi^+$ reaction,
the harmonic oscillator constant was obtained,
corresponding to a point-proton charge radius of $1.85(8)\,\mathrm{fm}$ for $^6\mathrm{He}_\text{g.s.}$.
However, their model could not reproduce the behaviour in the $\alpha+d\to{}^6\mathrm{Li}(3.563)+\pi^0$ reaction;
the exceptionally large cross section, together with a small asymmetry,
at $1.2\,\mathrm{MeV}$ above the threshold could not be reproduced.
In Ref.~\cite{Andersson2006}, the possibility of 
``a second source of $^6\mathrm{Li}$ ions with (or mimicking) a more symmetric
distribution in the c.m. frame'' was pointed out.
Nevertheless, any candidate reactions could not be identified.

Now, we suppose that some resonance, which can decay into $^6\mathrm{Li}(3.563)+\pi^0$,
exists near the $^6\mathrm{Li}(3.563)+\pi^0$ threshold. 
If the decay width of the resonance is sufficiently narrow, 
the formation of the resonance may 
explain the sharp increase and decrease in the cross section of 
the $\alpha+d \to {}^6\mathrm{Li}(3.563)+\pi^0$ reaction.
The small asymmetry may be qualitatively accounted for. 
Herein, the $\pi NN\alpha$ system described in Section 3 is proposed as a promising candidate for it,
while the theoretical calculations for the structure of this system and
its formation cross section in deuteron-alpha fusion are called for.

\section{\unboldmath Pion-Transfer Reaction to Populate the $\pi NN\alpha$ Resonance}
Currently, we cannot draw any strong conclusion from Fig.~\ref{CELSIUS}; however,
the intriguing result is worth considering.
A straightforward approach to validate the sharp structure of the excitation function shown in Fig.~\ref{CELSIUS} is to repeat 
the measurement of the pionic fusion reaction with finer steps in the beam energy.

Alternatively, we propose the $(p,d)$ reaction on a $^7\mathrm{Li}$ target
to populate the $\pi NN\alpha$ resonance.
Although the spectroscopy using the $(d,^3\mathrm{He})$ reaction
to transfer a $\pi^-$ to a target nucleus
is well established
in the study of deeply bound pionic states in heavy nuclei~\cite{Yamazaki2012,Nishi2018},
a similar pion-transfer reaction ($pn\to d\pi^0$) is desirable to populate 
the $\pi NN\alpha$ resonance via the $^6\mathrm{Li}(3.563)\text{-}\pi^0$ doorway from $^7\mathrm{Li}$.
Although the recoilless condition is satisfied
when the kinetic energy of the incident proton is approximately $330\,\mathrm{MeV}$,
a higher kinetic energy as well as scattering at finite angles,
both of which contribute to increasing the momentum transfer, are preferred for two reasons. 
First, an angular momentum transfer of $|\Delta \ell|=1$ is necessary
because a neutron in the $0p_{3/2}$ orbital is picked up and
a pion in the $s$ state is transferred to the neutron-hole state corresponding to $^6\mathrm{Li}(3.563)$.
Second, measurement of deuterons ejected at zero or small angles
with a momentum close to that of the beam protons 
is difficult because of a strong background
due to the non-interacting and elastically scattered protons.

We propose an experiment to perform a measurement using this reaction at Research Center for Nuclear Physics (RCNP), Osaka University~\cite{RCNP_E503}, which was already approved by the RCNP Beam-time Program Advisory Committee in March 2017.  
The incident beam energy will be $350$ or $392\,\mathrm{MeV}$.
A newly developed GRAF beamline~\cite{Kobayashi2019} will be used
to separate the ejectiles scattered at $4.5^\circ$ from the unreacted beam,
which is guided to a dump located in a shield wall around $25\,\mathrm{m}$ downstream of the target.
In these kinematical conditions, the momentum transfer is less than $100\,\mathrm{MeV}/c$.
Owing to the Grand Raiden high-resolution magnetic spectrometer as well as a high-intensity primary proton beam with a small energy spread of the order of $100\,\mathrm{keV}$,
a missing-mass spectrum for the $^7\mathrm{Li}(p,d)$ reaction near the $^6\mathrm{Li}(3.563)+\pi^0$ threshold
can be investigated precisely.
To address the experimental feasibility, the differential cross sections of not only the resonance formation but also the background are evaluated. In the following the beam energy is fixed to $392\,\mathrm{MeV}$ for simplicity. 

The differential cross section of the $^7\mathrm{Li}(p,d)$ reaction is estimated by multiplying 
the elementary cross section of $pn\to d\pi^0$ by the effective number of neutrons in the $0p_{3/2}$ orbital for the reaction. The spectroscopic factor for $^6\mathrm{Li}(3.563)$ is also taken into consideration to select the final state of $^6\mathrm{Li}(3.563)+\pi^0$.
\begin{itemize}
    \item The elementary cross section $(d\sigma/d\Omega)_{pn\to d\pi^0}$ is estimated to be $4.3\,\mathrm{mb/sr}$, using an empirical expression for that of the $pp\to d\pi^+$ reaction $(d\sigma/d\Omega)_{pp\to d\pi^+}$~\cite{Toki1991}. Here, isospin symmetry is assumed, leading to the relationship of $(d\sigma/d\Omega)_{pn\to d\pi^0}=(d\sigma/d\Omega)_{pp\to d\pi^+}/2$.
    \item Two neutrons occupy the $0p_{3/2}$ orbital in $^7\mathrm{Li}$. However, the effective number of neutrons in the reaction is reduced
    because of the initial and final state interaction for incoming protons and outgoing deuterons in the $^7\mathrm{Li}(p,d)$ reaction.
    It is estimated to be 0.16 in the Glauber model, using the total cross section of $p+N$ and $d+N$ reactions at relevant energies.
    \item The spectroscopic factor for neutron pick-up leading to $^6\mathrm{Li}(3.563)$ is assumed to be 0.24, according to a measurement of the $^7\mathrm{Li}(p,d)$ reaction at $185\,\mathrm{MeV}$~\cite{Fagerstrom1976}.
\end{itemize}
Under these assumptions, the differential cross section of $^7\mathrm{Li}(p,d){}^6\mathrm{Li}(3.563)+\pi^0$ is estimated to be $\approx 80\,\mathrm{\mu b/sr}$. If the sticking probability to form the $\pi NN\alpha$ resonance is 1\%, its formation cross section will be $\approx 0.8\,\mathrm{\mu b/sr}$.

On the other hand, the background spectrum can be inferred from that of $^{208}\mathrm{Pb}(n,d)$ reaction with the incident energy of $400\,\mathrm{MeV}$~\cite{Yamazaki1993}.
Supposing that the cross section is proportional to $A^{2/3}$, where $A$ is the mass number, the background level for the $^7\mathrm{Li}(p,d)$ reaction near the $\pi$ emission threshold will be of the order of $30\,\mathrm{\mu b/sr/MeV}$.
It should be noted that this estimation is in good agreement with the double differential cross section at $20^\circ$ for the $^{12}\mathrm{C}(p,d)$ reaction at $392\,\mathrm{MeV}$ measured at RCNP, which is around $20\text{--}30\,\mathrm{\mu b/sr/MeV}$~\cite{Uozumi2011}.

Using these estimations, it can be seen that the signal-to-noise ratio in the missing-mass spectrum may be of the order of $10^{-2}$ if the width of the $\pi NN\alpha$ peak is narrow enough as indicated by Fig.~\ref{CELSIUS}.
For example, when a proton beam with an intensity of $10\,\mathrm{nb}$ impinges on a $100\,\mathrm{mg/cm^2}$-thick $^7\mathrm{Li}$ target, thousands of $\pi NN\alpha$ resonances may be produced per hour, to be compared with the background level of $\approx 2\times 10^5/\mathrm{MeV}$ per hour.
Therefore, a high-resolution, high-statistics measurement will enable us to search for a signature of the $\pi NN\alpha$ resonance in the missing-mass spectrum. Even if a peak structure is not observed experimentally, the spectrum will provide unique information on the $\pi^0\text{-}{}^6\mathrm{Li}^{(*)}$ interaction.
\section{Conclusion}

For the first time, it is pointed out that the possibility that the sharp increase and decrease in the cross section of a pionic-fusion reaction, 
$\alpha+d\to{}^6\mathrm{Li}(3.563)+\pi^0$, observed in an experiment at CELSIUS, 
may be due to the formation of a $\pi NN\alpha$ resonance.
As a subsystem of $\pi NN$ is expected to have a resonance slightly above the threshold,
a four-body calculation of the structure of the $\pi NN\alpha$ system is worthwhile in elucidating the origin of the peculiar excitation function.
We also discussed the experimental feasibility of an investigation of the resonance
using the $^7\mathrm{Li}(p,d)$ reaction in terms of the formation cross section.

\ack
We would like to thank S.~Hirenzaki, E.~Hiyama, A.~Hosaka and K.~Itahashi for the helpful discussions.

\let\doi\relax

\end{document}